\documentclass[reprint,aps,prb,superscriptaddress,showpacs,longbibliography,floatfix]{revtex4-1}
\usepackage{units}
\usepackage{subfigure}
\usepackage{xspace,textcomp}
\pdfoutput=1
\usepackage{amsbsy,amssymb,amsmath}
\usepackage{graphicx,color,epsfig,rotate}
\usepackage{upgreek}
\usepackage{hyperref}% add hypertext capabilities
\newcommand{\ceossb}{CeOs\texorpdfstring{$_4$}{4}Sb\texorpdfstring{$_{12}$}{12}\xspace}

\begin{document}
\preprint{APS/123-QED}
\title{Unusual phase boundary of the magnetic-field-tuned valence transition 
in CeOs\texorpdfstring{$_4$}{4}Sb\texorpdfstring{$_{12}$}{12}}

\author{K. G\"{o}tze}
\author{M. J. Pearce}
\author{P. A. Goddard}
\affiliation{Department of Physics, University of Warwick, Coventry CV4 7AL, UK.}

\author{M. Jaime}
\affiliation{National High Magnetic Field Laboratory, Los Alamos National Laboratory, MS-E536, Los Alamos, New Mexico 87545, USA.}

\author{M. B. Maple}
\author{K. Sasmal}
\affiliation{Department of Physics, University of California, San Diego, La Jolla, CA 92093, USA.}

\author{T. Yanagisawa}
\affiliation{Department of Physics, Hokkaido University, Sapporo 060-0810, Japan}

\author{A. McCollam}
\author{T. Khouri}
\affiliation{High Field Magnet Laboratory (HFML-EMFL), Radboud University, Toernooiveld 7, 6525 ED,
Nijmegen, The Netherlands}

\author{P.-C. Ho}
\affiliation{Department of Physics, California State University, Fresno, CA 93740, USA.}

\author{J. Singleton}
\affiliation{National High Magnetic Field Laboratory, Los Alamos National Laboratory, MS-E536, Los Alamos, New Mexico 87545, USA.}

\date{\today}

\begin{abstract}

The phase diagram of the filled skutterudite \ceossb has been mapped in fields $H$ of up to \unit[60]{T} and temperatures $T$ down to \unit[0.5]{K} using resistivity, magnetostriction, and MHz conductivity. The valence transition separating the semimetallic low-$H$, low-$T$ $\cal{L}$ phase from the metallic high-$H$, high-$T$ $\cal{H}$ phase exhibits a very unusual, wedge-shaped phase boundary, with a non-monotonic gradient alternating between positive and negative. 
The expected ``elliptical'' behavior of the phase boundary of a valence transition with $H^2 \propto T^2$ originates in the $H$ and $T$ dependence of the free energy of the $f$~multiplet. Here, quantum oscillation measurements suggest that additional energy scales associated with a quantum critical point are responsible for the deviation of the phase boundary of \ceossb from this text-book behavior at high $H$ and low $T$. The distortion of the low-$H$, high-$T$ portion of the phase boundary may be associated with the proximity of \ceossb to a topological semimetal phase induced by uniaxial stress.

\end{abstract}

\maketitle
\section{Introduction}

\begin{figure}[t]
\begin{center}
		{\includegraphics[width=.98\columnwidth]{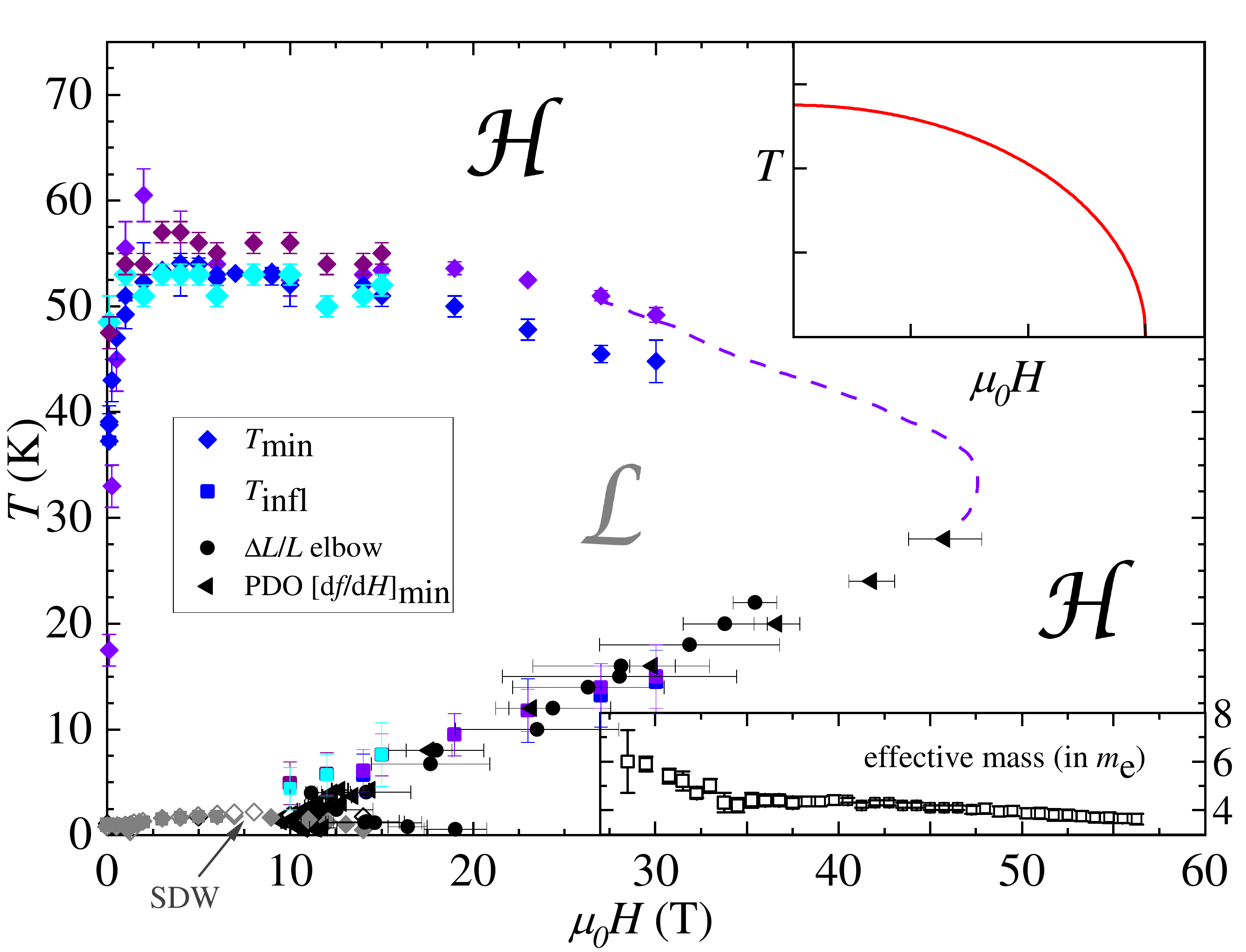}}
		\end{center}
		\vspace{-8mm}
   \caption[Phasediagram of \ceossb.]{$T$-$H$-phase diagram of 
   \ceossb derived from data in this work. Points are from
   resistivity (solid diamonds: $T_\mathrm{min}$, solid squares: 
   $T_\mathrm{infl}$, 
   different colors represent different samples), 
   magnetostriction (solid circles) and MHz conductivity 
   (solid triangles); SDW phase boundary from 
   Refs.~\cite{namiki_2003,sugawara_2005,rotundu_2006,tayama_2015} (grey symbols).
   The dotted line is a guide to the eye of a plausible $\cal{L}$-$\cal{H}$ phase boundary completion.
    Upper inset: example for the ``elliptical'' $H^2 \propto T^2$ phase boundary expected for valence transitions. 
   Lower inset: field dependence of the effective 
   mass.}
   \vspace{-5.5mm}
	\label{fig:COS_phasediagr}
\end{figure}

Valence transitions, in which $f$-electrons 
undergo a temperature and/or magnetic-field driven 
transformation from itinerant to quasi-localized, are 
associated with significant 
changes in material properties~\cite{ho_2016,gschneider,drymiotis_2005} 
and dramatic alterations 
to the Fermi surface \cite{ho_2016,dzero_2000}.
Perhaps the best-known is the $\gamma-\alpha$ 
transition in Ce and its alloys, 
which leads to a spectacular sample volume collapse \cite{gschneider,drymiotis_2005}. 
Valence transitions are also thought to be responsible 
for the onset of the ``hidden order phase'' of 
URu$_2$Si$_2$, plus some phase boundaries of elemental Pu~\cite{harrison_2019_URS,lashley_2003,Harrison_2019_Pu}. 
A key identifying feature of valence transitions is 
the resulting {\it elliptical phase boundary} in which the critical
magnetic field $H$ and temperature 
$T$ of the valence transition 
follow a $H^2 \propto T^2$ behavior (upper inset of Fig.~\ref{fig:COS_phasediagr}) or, in other words, 
lie on a straight 
line when plotted as $H^2$ versus $T^2$, with the slope 
determined by the $g$-factor alone \cite{dzero_2000}.

By contrast, we show here that the valence 
transition in \ceossb, 
identified by its effect
on the Fermi surface and material 
properties~\cite{ho_2016}, does not follow
the above textbook elliptical behavior.
We performed MHz conductivity, 
magnetostriction and resistivity measurements on \ceossb, to
map out the phase diagram shown in 
Fig~\ref{fig:COS_phasediagr}. 
The valence transition separates the low-$T$, 
low-$H$, semimetallic $\cal{L}$ phase and 
the high-$T$, high-$H$, metallic $\cal{H}$ phase~\cite{ho_2016};
it is immediately obvious that it behaves very unconventionally, 
falling back to lower $T$ as $H \rightarrow 0$ 
and lower $H$ as $T\rightarrow 0$.
We suggest this unusual behavior is due to sensitivity of 
the groundstates of \ceossb
to quantum fluctuations, and proximity to a topological semimetallic phase.

\ceossb is part of an interesting series of rare-earth-based 
filled skutterudites
including the unconventional superconductor PrOs$_4$Sb$_{12}$~\cite{bauer_2002,maple_2002} 
and ferromagnetic NdOs$_4$Sb$_{12}$ \cite{ho_2016,ho_2005}. 
While PrOs$_4$Sb$_{12}$ and NdOs$_4$Sb$_{12}$ possess 
similar Fermi surfaces comprising 
multiple pockets (but different effective masses), 
\ceossb exhibits a valence transition from the heavy-effective-mass
(Sommerfeld coefficient $\gamma=\unit[92]{mJ~mol^{-1}~K^{-2}}$) semimetallic 
$\cal{L}-$phase~\cite{bauer_2001,sugawara_2005,harima_2003,yan_2012}
to the $\cal{H}-$phase, characterized by a simple, 
almost spherical Fermi surface with a light 
effective mass~\cite{ho_2016}.

Earlier studies \cite{bauer_2001} suggested \ceossb to be a Kondo insulator due to the resistivity increase at low temperatures but
band-structure calculations for the $\cal{L}$-phase 
have confirmed the semimetallic, gapless groundstate with heavy masses under ambient conditions
and predict the system to become a topological semimetal or 
topological Kondo insulator under applied strain~\cite{harima_2003,yan_2012}.

Antiferromagnetic order, believed to be due to 
spin-density 
wave (SDW) formation, was observed in \ceossb
below \unit[1]{K} at $H=0$~\cite{namiki_2003,yogi_2005,yang_2006}.
The transition temperature, $T_{\rm SDW}$,
was seen to increase with increasing field to \unit[2]{K} at 
\unit[7]{T} and subsequently decrease \cite{ho_2016,sugawara_2005,rotundu_2006} and be suppressed around 15~T.

In Ref.~\cite{ho_2016}, an attempt to trace the 
high-temperature limits of the $\cal{L-H}$ boundary
was made using $\chi = \partial M/\partial H$ contours, where $M$
is the magnetization. However, subsequent
experiments suggested that $\chi$ was not an accurate indicator of the valence transition, 
prompting the current comprehensive series of measurements
that reveal the much more unusual behavior
shown in Fig.~\ref{fig:COS_phasediagr}.

Our choice of techniques for clarifying the phase boundary is based
on previous experimental evidence of the valence transition
in cerium-based systems. The strongest indicator of this valence transition 
is the drastic change in unit cell volume.
Magnetostriction measurements indicating
the change in sample length as a function
of field can very sensitively detect such
volume changes and allow for a precise determination
of the critical field of the phase transition. 
Due to the ratio of 1:17 of cerium to other elements in \ceossb we expect a smooth rather than a sharp structural transition as in elemental or slightly doped cerium \cite{gschneider,drymiotis_2005}.

Electrical resistivity (and related methods
like MHz conductivity measurements) have been shown
to be another reliable indicator of the valence transition
in cerium and its alloys: hysteretic behavior in resistivity
as a function of temperature or pressure
in Ce$_{0.8}$La$_{0.1}$Th$_{0.1}$ \cite{thompson_1983,drymiotis_2005},
or CeNi \cite{gignoux_1985} occurs at the valence transition
due to inhomogeneous strain fields in the sample
associated with the cell volume collapse.

\section{Experimental details}

\ceossb single crystals were prepared using a 
molten-flux technique \cite{bauer_2001} 
with Sb excess (details can be found in \cite{ho_2016}). 
Several single crystals (cubic space group  $\mathrm{T}_\mathrm{h}^5$ (Im$\overline{3}$), No.~204)
were obtained from the same growth batch.
Four bar-shaped crystals (samples B1--B4) were 
used to measure standard four-probe resistivity with the current 
applied along the [100] axis.

Magnetic fields up to \unit[15]{T} (\unit[30]{T}) were provided by superconducting 
(water-cooled resistive) magnets.
A Proximity-Detector-Oscillator (PDO) technique
was used for contactless (MHz) resistivity measurements in pulsed magnetic fields. Shifts in the PDO frequency $f$
are caused by alterations in 
the sample skin depth~\cite{ghannadzadeh_2011,altarawneh_2009},
leading to $\Delta f \propto -\Delta \rho$  for 
small relative changes in $\rho$~\cite{ghannadzadeh_2011}.
Magnetostriction was measured in pulsed fields by the Fiber Bragg Grating technique~\cite{jaime_2017}. The magnetic field was applied along [001] for all measurements.

\section{Results}
\subsection{Resistivity - Temperature-sweeps at fixed field}

\begin{figure*}[htbp]
%\vspace{-35mm}
\includegraphics[width=0.99\textwidth]{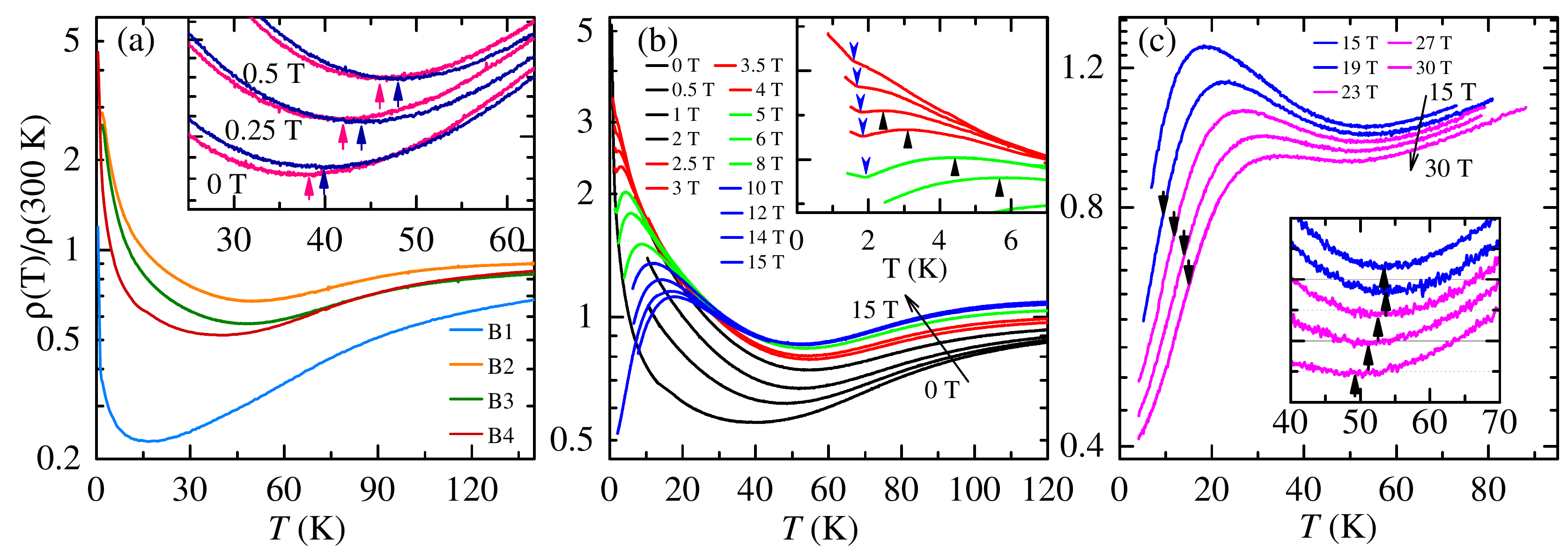}
   \vspace{-3mm}
   \caption[$\rho$(T) of sample B4 for several magnetic fields and derived phase diagram.]{
	(a) Resistivity $\rho$ [normalized to $\rho(300~{\rm K}$)] of all four measured crystals (B1--B4) at $H=0$ for $0.6 \leq T \leq 140$~K. 
	Inset: hysteresis between warming (red) and cooling (blue) for sample B4 at low fields. Arrows mark the minimum.
	(b) Normalized $\rho(T)$ of sample~B4 for several fields $\leq 15$~T. An arrow at high temperatures indicates increasing field; note that the resistivity maximum at low temperatures decreases with increasing field. Inset: low-$T$ behavior of $\rho (T)$ between 2.5 and \unit[6]{T}. 
	Up arrows mark the emerging maximum for $\mu_0H \geq 3$~T, down arrows mark the SDW transition.
	(c) Resistivity of sample B1 for $15 \leq \mu_0H \leq 30$~T. Curves are offset for clarity. Black arrows mark the inflection point for $\mu_0 H >\unit[15]{T}$. Inset: $\rho$ minimum; arrows track the suppression of $T_\mathrm{min}$ with increasing field.
	}
	\label{fig:COS_RvsT}
	\vspace{-4mm}
\end{figure*}

Fig.~\ref{fig:COS_RvsT}\,(a) shows $\rho(T)$
for crystals B1--B4 at zero field. 
In agreement with previous studies~\cite{bauer_2001,sugawara_2005},
resistivity initially decreases upon cooling, reaching a minimum 
at $T = T_{\rm min}$.
Below this temperature, $\rho(T)$ increases strongly.
The minimum in $\rho$ has previously been interpreted as 
the $\cal{H-L}$ transition~\cite{ho_2016}. In contrast to
that earlier work, our study
shows that the exact value of 
$T_\mathrm{min}$ at $H=0$ is strongly sample-dependent,
ranging from 17.5~K to 48.5~K for the four crystals measured. 
Similar sample-dependent variations 
in the low-temperature magnetic and transport 
properties of other Ce-based skutterudites were observed 
in~\cite{meisner_1985} and \cite{bauer_cerusb_2001}, 
and attributed to a delicate balance 
between competing scattering effects and the 
influence of magnetic impurities. A more detailed discussion on the sample dependence in our measurements is provided in section~\ref{sec:sampledep}.

Note that there is a small but consistent 
hysteresis between cooling and 
warming through the $\rho(T)$ minimum, 
with $T_\mathrm{min}$ 
being higher on cooling than on warming [inset of Fig.~\ref{fig:COS_RvsT}\,(a)]. 
The difference $\Delta T$ increases with temperature-sweep rate, 
but it is always non-zero even for the slowest temperature changes. 
With decreasing temperature-sweep rate $\Delta T$ converges to $\approx \unit[0.5]{K}$
indicating that it originates from a phase transition rather than a lag in thermal equilibrium.

As mentioned in the introduction, 
similar hysteretic behavior in $\rho$
was observed 
close to the valence transition in Ce-based materials \cite{thompson_1983,gignoux_1985,drymiotis_2005}
but also  
close to the valence transitions in 
YbInCu$_4$~\cite{immer_1997}, 
supporting the proposal~\cite{ho_2016} that the valence of 
Ce in \ceossb changes at $T_\mathrm{min}$.

The shape of the $\rho(T)$ curves 
changes significantly when a magnetic field is applied. 
Fig.~\ref{fig:COS_RvsT}\,(b) shows normalized 
resistivity curves $\rho(T)/\rho(\unit[300]{K})$ for 
sample B4 for fields up to \unit[15]{T}. 
Increasing the magnetic field initially shifts the position of the minimum to higher temperatures; subsequently, 
$T_\mathrm{min}$ is almost field independent
between 3 and \unit[15]{T}, 
but then decreases [inset of Fig.~\ref{fig:COS_RvsT}\,(c)] for higher fields.

The transition to the ordered SDW phase manifests itself by a the kink in the $\rho(T)$ curve, marked by down arrows in the
inset of Fig.~\ref{fig:COS_RvsT}\,(b). Similar features were observed at the transition to the SDW in earlier
measurements \cite{sugawara_2005},
and the behavior of $T_\mathrm{SDW}$ in our data
accords with the results in~\cite{sugawara_2005}.
 
For $\mu_0H >\unit[3]{T}$, a local $\rho(T)$ maximum
develops at temperature $T_{\rm max}$
just above $T_\mathrm{SDW}$ [up arrows in 
the inset of Fig.~\ref{fig:COS_RvsT}\,(b)] and moves higher with 
increasing field. 
$T_\mathrm{max}$ shows an almost linear field dependence (open diamonds in Fig.~\ref{fig:COS_logphasediagr});
in addition, as $H$ grows, the $\rho$($T$) 
maximum becomes broader 
and lower. These trends continue 
for fields up to \unit[30]{T}  
[Fig.~\ref{fig:COS_RvsT}\,(c)]. 

As will be clear from Fig.~\ref{fig:COS_phasediagr},
for fields higher than 10~T, it is possible for a temperature sweep at constant field to traverse the valence transition twice:
$\cal{H-L}$, followed by $\cal{L-H}$.
In this context, the $\rho(T)$ maximum at $T_{\rm max}$ is a precursor that occurs before
the restoration of 
metallic behavior at low $T$ and high $H$ but it does not indicate a phase transition.
No hysteresis was observed around the maximum
supporting the interpretation that the valence does not change at $T_{\rm max}$.
For $\mu_0 H \geq\unit[10]{T}$, we find a $\rho(T)$ inflection 
point at $T=T_{\rm infl}$ below which 
metallic resistivity $\rho$($T$)=$\rho_0+AT^2$ is obeyed. 
It is this inflection point that 
we attribute to the 
valence transition. 
$T_{\rm infl}$ is marked by down arrows 
in Fig.~\ref{fig:COS_RvsT}\,(c) and indicated by solid squares in Figs.~\ref{fig:COS_phasediagr} and \ref{fig:COS_logphasediagr}.
For fields between \unit[3.5]{T}, where the $\rho(T)$ maximum
first emerges, 
and $\sim\unit[9]{T}$, the SDW phase interposes itself
and $T_{\rm infl}$ is not visible.
Nevertheless, within the region 
between $T_\mathrm{max}$ and $T_\mathrm{SDW}$, 
\ceossb begins to revert 
to metallic behavior ({\it i.e.,} $\rho(T)$
falls as $T$ decreases).

\subsection{MHz Resistivity and Magnetostriction at fixed temperatures in pulsed magnetic fields}

\begin{figure*}[htbp]
%\vspace{-39mm}
\includegraphics[width=0.99\textwidth]{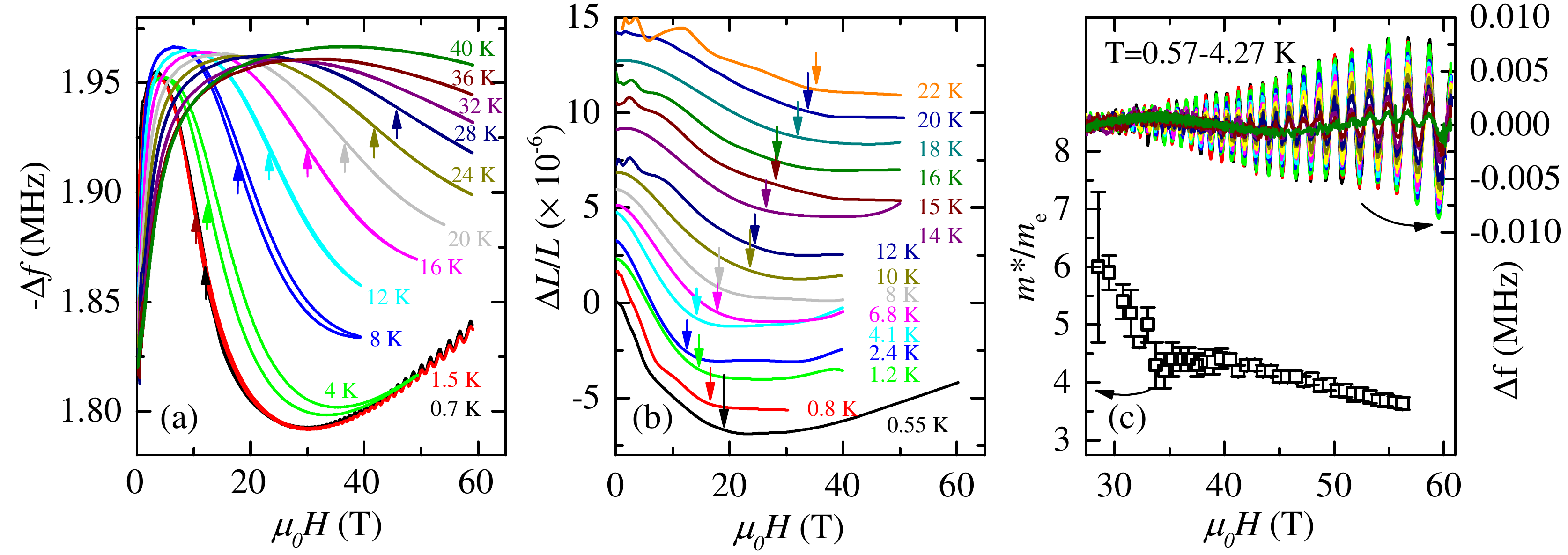}
   \vspace{-3mm}
   \caption[PDO results, pulsed field magnetostriction, and field dependent effective mass.]{(a) PDO frequency change $-\Delta f$ in pulsed fields. Note the hysteresis between up and down sweeps.
   (b) Pulsed-field magnetostriction $\Delta L/L$
   (Data offset for clarity). Arrows mark the valence transition in (a) and (b).
		(c) Oscillatory part of $-\Delta f$ for $0.57 \leq T \leq \unit[4.27]{K}$, and field-dependence of the effective mass of the $F=\unit[1.6]{kT}$ quantum oscillations; $m_\mathrm{e}$ is the bare electron mass.}
   \vspace{-5mm}
\label{fig:COS_highfield}
\end{figure*}

We will now turn to the high-$H$, low-$T$ part
of the $\cal{L-H}$ phase boundary that partially coincides
with the SDW phase suppression. 
It was initially identified using PDO 
experiments for $T \leq 4$~K~\cite{ho_2016}.
Here,
we use the same technique to track the transition to higher temperatures.
Fig.~\ref{fig:COS_highfield}\,(a) shows the PDO frequency change
in \ceossb in fields up 
to \unit[60]{T} and temperatures between 0.7 and \unit[40]{K}.
For low temperatures a pronounced maximum exists at low fields,
followed by a sharp decrease of $-\Delta f$ and a 
minimum at high fields.   
Since $-\Delta f$ is proportional to $\Delta \rho$, 
we can analyse the PDO $-\Delta f$ data in an analogous manner
to the $\rho(T)$ data.
The maxima in both properties have the same field
dependence, and the PDO maximum continues to move linearly 
to higher fields with increasing temperature. 
As in the $\rho(T)$ data, the maximum is the 
precursor of a change from semimetallic 
behavior ($\cal{L}$ phase) to metallic 
character ($\cal{H}$ phase) and not a phase transition.
The succeeding drop in $\rho$ (or $-\Delta f$) is 
commensurate with
removing Ce $f$-electrons from the 
$\cal{L-}$semimetallic groundstate with its 
ultra-heavy effective masses, 
the resulting Fermi energy shift
producing the larger, almost-spherical 
Fermi surface (with light-mass
quasiparticles) of the $\cal{H-}$phase predicted
by theory and observed in experiment \cite{harima_2003,yan_2012,ho_2016}. 
As in Ref.~\cite{ho_2016}, we identify the $\cal{L-H}$ 
transition as the inflection point within
the fall in $-\Delta f$ [arrows in
Fig.~\ref{fig:COS_highfield}\,(a)].
Hysteresis occurs between PDO data recorded 
with rising 
and falling field, again suggestive of the lossy kinetics typical
of valence transitions \cite{drymiotis_2005}.

Magnetostriction measurements 
were carried out in pulsed magnetic fields up to 60~T and for temperatures between 
0.5 and \unit[40]{K}
in order to track sample volume changes that accompany the valence transition
\cite{gschneider,dzero_2000,drymiotis_2005}.
Typical results are shown in Fig.~\ref{fig:COS_highfield}\,(b)
as fractional change in length 
($\Delta L/L$) versus field. 
The $\cal{L-H}$ phase transition is marked by a change of slope: 
the initial decrease of $\Delta L/L$ slows down and is reversed
or becomes flat, causing an elbow in the data.
Note that only one in 17 atoms in \ceossb is 
cerium which is expected to lead to a smoother structural transition
compared to elemental or slightly doped cerium \cite{gschneider,drymiotis_2005}.

Linear functions are fitted to the data below 
and above the elbow;
the transition field is defined as the point 
at which the gradient of the data 
is equal to the mean gradient of
the two linear functions and is indicated by arrows in the graph. Examples of raw and averaged data are shown in the Supplementary Information, section~S1.
At low temperatures, the valence transition follows the established SDW 
border to lower $H$ as $T$ increases \cite{namiki_2003,sugawara_2005,rotundu_2006,tayama_2015}. 
Above \unit[4]{K} this trend reverses; the transition field starts 
to grow with $T$.
The elbow can be followed up to \unit[22]{K}; at higher temperatures,
it is too weak to be identified reliably.

\subsection{Quantum oscillations}

Shubnikov-de Haas oscillations in $-\Delta f$ 
occur in the lower-$T$ curves in 
Fig.~\ref{fig:COS_highfield}\,(a) above \unit[25]{T}. 
These oscillations comprise a single frequency $F \approx \unit[1600]{T}$, 
due to the roughly 
spherical $\cal{H-}$phase Fermi surface~\cite{ho_2016}.
Using an analysis similar to Ref.~\onlinecite{rebar},
the quasiparticle effective mass 
$(m^*)$ was found to be field dependent. Details and examples can be found in the Supplementary Information, section~S2.
Fig.~\ref{fig:COS_highfield}\,(c) shows the oscillating 
part of $-\Delta f$ (right axis) and the development 
of $m{^*}$ for several mean fields $B_{\rm m}$ 
separated by \unit[0.75]{T} steps.  
At first, $m^*$ increases slowly 
with decreasing field, from 3.6 to \unit[4.4]{$m_{\rm e}$} 
between 56 and \unit[35]{T}. 
As the intersection of the 
$\cal{L}$, $\cal{H}$ and SDW phases at lower field
approaches, $m^*$ increases rapidly, 
reaching \unit[6]{$m_{\rm e}$} at \unit[28.5]{T}, 
the lowest field at which a value could be determined.

Quantum oscillations in the $\cal{L}$-phase have not been observed experimentally. The cross-sections of the calculated Fermi surfaces are quite small ($\approx 100$ times smaller compared to the $\cal{H}$-phase; see Supplementary Information, section~S3, for more information) and the quasiparticles possess heavy masses \cite{bauer_2001,harima_2003,yan_2012}. Very low temperatures would therefore be required to observe quantum oscillations from heavy quasiparticles. However, the presence of the ordered SDW phase below 2~K would prevent their observation in this temperature regime because of the accompanying Fermi surface reconstruction. 

A discussion on whether the field-tuned $\cal{L-H}$ transition might be affected by the quantum limit of quantum oscillations in the $\cal{L}$ phase can be found in the Supplementary Information, section~S3.

\section{Discussion}
\label{sec:disc}

\subsection{Phase diagram}

\begin{figure}[t]
\begin{center}
		{\includegraphics[width=.95\columnwidth]{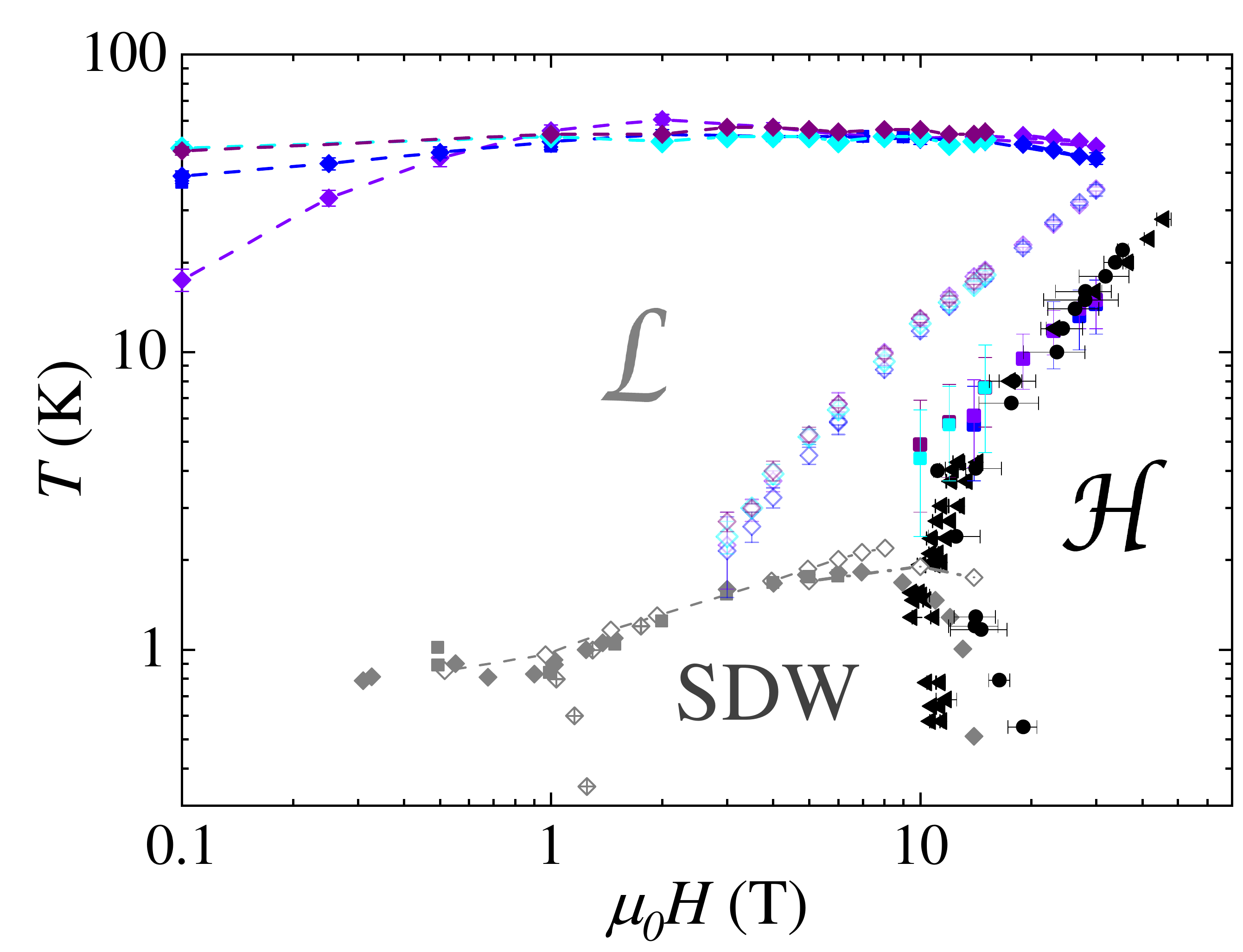}}
		\end{center}
   \caption[Logaritmic phasediagram of \ceossb.]{$T$-$H$-phase diagram of 
   \ceossb  with logarithmically scaled axes.
		Points are the same as in Fig.~\ref{fig:COS_phasediagr} with the addition of $T_\mathrm{max}$ (open diamonds).   
	 $T_\mathrm{max}$ does not mark a phase transition, but the precursor of the return to metallic behavior at low $T$/high $H$.
    }
	\label{fig:COS_logphasediagr}
\end{figure}

The PDO, magnetostriction and transport
data ($T_\mathrm{min}$
and $T_\mathrm{infl}$) for B1--B4
are shown in Fig.~\ref{fig:COS_phasediagr} with linear axes and in Fig.~\ref{fig:COS_logphasediagr} in logarithmic scaling.
The SDW phase boundaries were taken from Refs.~\cite{namiki_2003,sugawara_2005,rotundu_2006,tayama_2015}; the high field data above 10~T corresponds closely to our measurements.
The position of the valence transition could be determined quite precisely and agrees for different techniques. Further indications for a thermodynamic phase transition are the hysteresis in transport and PDO, and the change in behaviour of the lattice indicated by magnetostriction.

It is obvious that the ``wedge-shaped'' phase boundary
surrounding the $\cal{L}$ phase is very unusual indeed:
On the high-$T$, low-$H$ side of the phase diagram, 
the $\cal{H-L}$ transition temperature
$T_\mathrm{min}$, as mentioned above, differs 
for different samples, 
ranging from 17.5 to \unit[48.5]{K} at $H=0$. 
Additionally, with increasing field for
$0 \leq \mu_0H \leq \unit[2]{T}$, 
the transition at first
moves rapidly to higher temperatures ({\it i.e.,} has a positive gradient).
Subsequently, $T_{\rm min}$ hardly changes between 
2 and \unit[15]{T} but eventually decreases
in higher fields. 
Amongst the different samples,
the difference between the 
$T_{\rm min}$ values 
decreases above 2~T, but remains at least 
\unit[4]{K} up to \unit[30]{T}.

On the low-$T$, high-$H$ side, the metallic $\cal{H}$ phase is restored.
The field-induced $\cal{L-H}$ transition 
and SDW phase destruction coincide 
below $T\approx \unit[2]{K}$ and above \unit[10]{T}.
A simple interpretation is
that the SDW formation
is dependent on details
of the $\cal{L-}$phase Fermi surface topology~\cite{singleton};
once the $\cal{L-}$phase is removed by
the valence transition, the SDW will inevitably
be destroyed.
However, we will see below that the death of the
SDW and the $\cal{L-H}$ phase boundary are,
chicken-and-egg-like,
much more subtly entwined than in this simple interpretation.
Whereas the low$-T$ part of
the high-field phase boundary has a (relatively conventional)
negative gradient,
above $T=\unit[2]{K}$, the phase boundary assumes
a positive gradient; the $\cal{L-H}$ transition 
moves toward higher field with increasing
temperature, showing an almost linear $H,T$ relationship up to \unit[28]{K}.

The local minimum in $\rho$($T$), indicating the $\cal{H-L}$ transition,
and the $\rho$($T$) maximum, a precursor to the $\cal{L-H}$ transition,
move closer for increasing field, resulting
in a plateau-like feature at \unit[30]{T} (see Fig.~\ref{fig:COS_RvsT}\,(c)).
This convergence shows that the temperature range for which 
the $\cal{L-}$phase is stable 
shrinks with increasing field.
Extrapolating the measured points
suggests that above $\sim\unit[55]{T}$, $\cal{H}$ 
should be the only stable phase of \ceossb. 
A logarithmic scaling of the phase diagram displays 
this behavior more clearly, as shown in Fig.~\ref{fig:COS_logphasediagr}\cite{fn}.

\subsection{Energy scales and quantum criticality}

The $\cal{L-H}$ phase boundary is clearly different from
most others encountered in condensed-matter physics;
unlike {\it e.g.,} a mean-field boundary \cite{CandL}
or the ``domes'' observed in many correlated-electron
systems such as organic, high-$T_{\rm c}$ and pnictide
superconductors~\cite{singleton,ramshaw,michon,pnictides}, 
the gradient of the boundary in
$T,H$ space does not change monotonically,
but alternates {\it positive, negative, positive, negative}.
By the same token, it clearly deviates from the 
{\it elliptic} $(H^2 \propto T^2)$
behavior found for field-induced valence 
transitions in other cerium~based systems \cite{gschneider,drymiotis_2005}
or uranium compounds \cite{harrison_2019_URS}. 

The elliptic phase boundary usually associated with
valence transitions is driven by the
$T$- and $H$-dependencies of the
free energy of the quasi-localized $f$ 
multiplet \cite{dzero_2000}. 
The energies
of the Fermi liquids on either 
side of the phase boundary
will depend only slightly on $H$ and $T$,
so the multiplet's free energy
dominates the situation and drives the valence transition.
The $-TS$ (where $S$ is entropy) term in the free energy
means the phase in which the multiplet
is populated will always be the groundstate at high $T$ and
high $H$~\cite{dzero_2000,drymiotis_2005}, 
the multiplet's simple partition function resulting 
in the $H^2 \propto T^2$
elliptic boundary~\cite{dzero_2000}.
The deviation of \ceossb from this simple
behavior implies that one or more additional
energy scales that depend strongly on
$H$ and/or $T$ are present.

Turning first to the low$-T$, high$-H$ portion
of the valence transition, recall that the
$\cal{H}$-phase quasiparticle effective mass
appears to diverge as the $\cal{H-L}$
transition approaches (Fig.~\ref{fig:COS_highfield}\,(c)).
The antiferromagnetic PrOs$_4$As$_{12}$ shows a similar mass
increase on approaching the phase boundary of its
magnetic groundstate~\cite{ho_2007}.
Such effective-mass increases are frequently
associated with proximity to a quantum-critical
point (QCP)~\cite{ramshaw,michon,ho_2007}.
In the case of \ceossb,
the QCP is most likely associated with 
the field-driven SDW collapse
(Figs.~\ref{fig:COS_phasediagr} and \ref{fig:COS_logphasediagr}).
As $T\rightarrow 0$, the entropy $S$ will diminish as well in accordance with the third law of thermodynamics, leaving any other energy scale to dominate \cite{blundellblundell}. 
In consequence, strong quantum fluctuations
-~probably antiferromagnetic~- around the
QCP will greatly perturb the Fermi liquids' free energy
on either side of 
the $\cal{L-H}$ boundary~\cite{fn2}, challenging the
dominance of the multiplet's $-TS$ contribution~\cite{dzero_2000}.

\subsection{Sample dependence}
\label{sec:sampledep}

Moving to the high$-T$, $H\rightarrow 0$
portion of the $\cal{L-H}$ transition,
the most striking feature is the initial,
large, positive gradient.
Qualitatively similar 
behavior is seen 
in the phase diagram of a reduced-dimensionality 
antiferromagnet~\cite{sengupta};
in that case, the effect is attributed to thermal 
fluctuations affecting the system's free energy.
However, the phase-boundary gradient reversal measured in
Ref.~\onlinecite{sengupta} is much less marked than
that in \ceossb (Fig.~\ref{fig:COS_phasediagr}). 
Moreover, there is no obvious
reason why the mechanism of Ref.~\cite{sengupta}
would yield the strong sample dependence seen here.

The $\cal{L}$ phase of \ceossb is thought to be 
highly unusual among Ce compounds in that very small
application of uniaxial stress can transform it
into an unusual topological semimetal \cite{yan_2012}.
It is therefore possible that the inclusions
(fraction of a \% level)
of elemental Ce or Os,
in otherwise very high quality crystals,
recently observed in
neutron scattering experiments \cite{neutrons}
result in local regions of
varying uniaxial stress inducing 
topologically-protected
``domains''.
The number of domains would likely
be very dependent on sample quality,
and even small variations in strain within 
the same batch could give rise to the observed sample dependence.
In addition, the band structure associated with such states 
can be sensitively dependent on
magnetic field (see {\it e.g.} Refs.~\cite{rebar,tang}
and citations therein) perhaps leading to the initial positive gradient of the $\cal{L}$-$\cal{H}$ boundary.
Experimentally, the presence of 
even a small fraction of similar quasiparticles
in an otherwise unremarkable Fermi liquid
can have a disproportionate
effect on measurable macroscopic properties \cite{rebar}.
Analogous effects may well impact
the unusual low$-H$, high$-T$
curvature of the \ceossb valence transition.

\section{Summary}

In summary, the $H$-$T$-phase diagram of \ceossb 
has been mapped using resistivity, magnetostriction, and MHz conductivity.
The semimetallic $\cal{L}-$phase and the metallic $\cal{H}$-phase are separated by a valence transition 
that exhibits a wedge-shaped phase boundary that is clearly 
distinct from the text-book ``elliptical'' phase
boundary usually followed by valence transitions.
Field-dependent effective masses revealed by Shubnikov-de Haas oscillations within the $\cal{H}-$phase 
show an increasing $m^*$ as the field 
drops toward the $\cal{H-L}$ phase boundary,
suggesting proximity to a QCP.
The associated magnetic fluctuations may be responsible for 
the anomalous $H,T$ dependence of the valence transition
at high field.
The unusual low$-H$, high$-T$
portion of the phase boundary
may in contrast be associated with the
proximity of \ceossb to
a topological semimetal
induced by uniaxial stress,
resulting in strongly sample-dependent
behavior.

\section*{acknowledgments}

We thank Roger Johnson for constructive discussions. We would also like to acknowledge useful discussions with Qimiao Si and Piers Coleman.

Work performed at the University of Warwick is supported by the EPSRC and European Research Council (ERC) under the European Union’s Horizon 2020 research and innovation programme (grant agreement No 681260); at the National High Magnetic Field Laboratory, USA, by NSF Cooperative Agreements DMR-1157490 and DMR-1644779, the State of Florida, U.S. DoE, and through the DoE Basic Energy Science Field Work Project Science in 100 T; at UCSD by NSF DMR-1206553 and US DOE DEFG02-04ER46105; at Hokkaido University by JSPS KAKENHI JP15KK0146, JP18H04297, JP17K05525; at CSU-Fresno NSF DMR-1506677. We acknowledge the support of the HFML-RU/FOM, member of the European Magnetic Field Laboratory (EMFL).

\bibliography{COS_transport_Bib}

\begin{thebibliography}{40}
\bibitem [{fn()}]{fn}%
  \BibitemOpen
  \href@noop {} {}\bibinfo {note} {The ${\cal L-H}$ transition becomes
  difficult to observe in pulsed-field PDO data above \unit[28]{K}, as it
  merges with the high-$T$ part of the phase boundary close to \unit[40]{K}. As
  the upper part of the boundary is very flat in the $T,H$ plot, it is
  challenging to resolve in any swept-field, constant-$T$ pulsed experiment;
  the corresponding feature in the data will be very broad. Swept-$T$,
  fixed-$H$ PDO measurements would be most suitable to follow the flatter,
  upper boundary to higher fields but would be limited to $\leq \unit[30]{T}$
  by the resistive magnet used.}\BibitemShut {Stop}%
\end{thebibliography}

\end{document}

\preprint{APS/123-QED}
\title{Supplemental Material to ``Unusual phase boundary of the magnetic-field-tuned valence transition 
in CeOs\texorpdfstring{$_4$}{4}Sb\texorpdfstring{$_{12}$}{12}''}

\author{K. G\"{o}tze}
\author{M. J. Pearce}
\author{P. A. Goddard}
\affiliation{Department of Physics, University of Warwick, Coventry CV4 7AL, UK.}

\author{M. Jaime}
\affiliation{National High Magnetic Field Laboratory, Los Alamos National Laboratory, MS-E536, Los Alamos, New Mexico 87545, USA.}

\author{M. B. Maple}
\author{K. Sasmal}
\affiliation{Department of Physics, University of California, San Diego, La Jolla, CA 92093, USA.}

\author{T. Yanagisawa}
\affiliation{Department of Physics, Hokkaido University, Sapporo 060-0810, Japan}

\author{A. McCollam}
\author{T. Khouri}
\affiliation{High Field Magnet Laboratory (HFML-EMFL), Radboud University, Toernooiveld 7, 6525 ED,
Nijmegen, The Netherlands}

\author{P.-C. Ho}
\affiliation{Department of Physics, California State University, Fresno, CA 93740, USA.}

\author{J. Singleton}
\affiliation{National High Magnetic Field Laboratory, Los Alamos National Laboratory, MS-E536, Los Alamos, New Mexico 87545, USA.}

\date{\today}

\maketitle

\section{Processing of Magnetostriction data}
\label{sec:magnetostr}

\begin{figure}
\begin{center}
		{\includegraphics[width=.99\columnwidth]{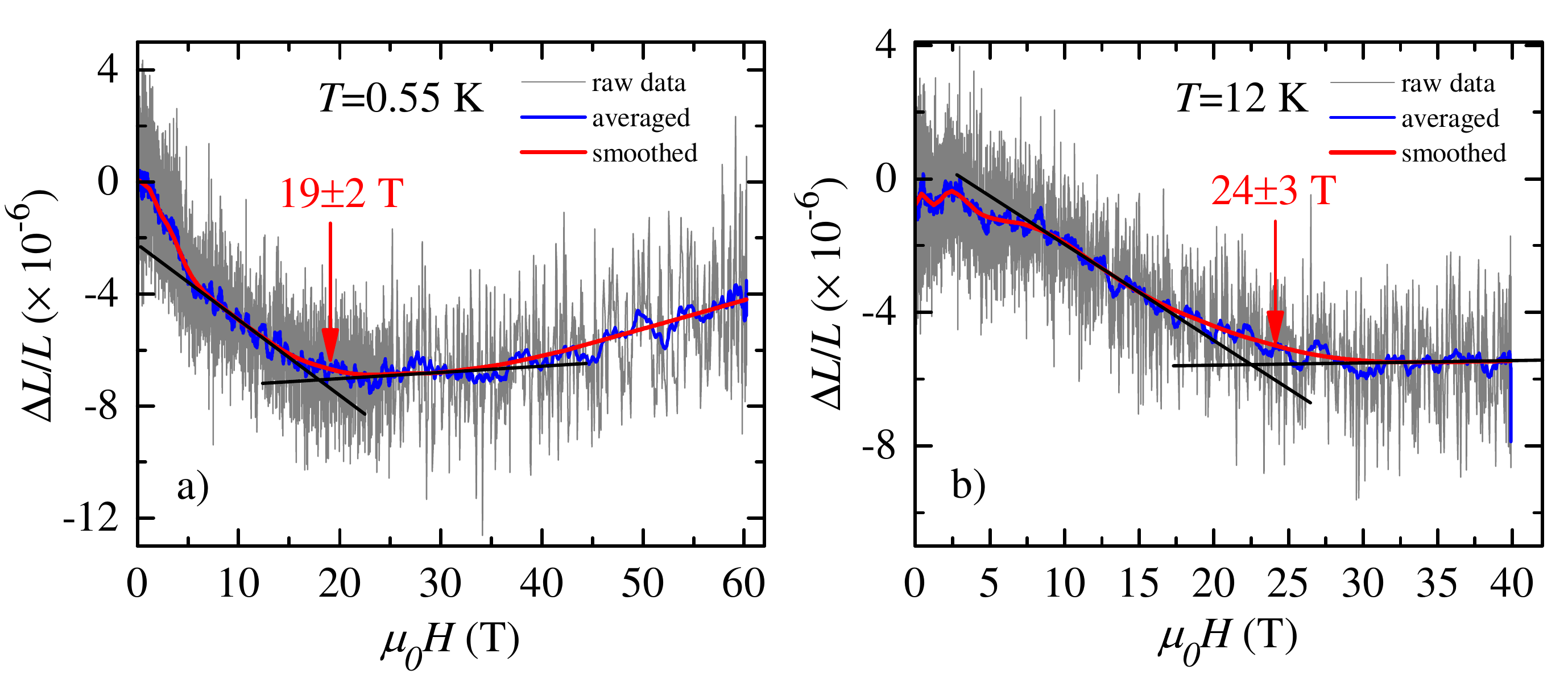}}
		\end{center}
   \caption{Raw (grey), averaged (blue) and smoothed (red) data from the magnetostriction experiment. Black lines: linear fits above and below the ``elbow'' explained below. (a) $T=\unit[0.55]{K}$, (b) $T=\unit[12]{K}$.}
	\label{fig:magnetostr}
\end{figure}

Fig.~\ref{fig:magnetostr} shows magnetostriction data for two temperatures: (a) 0.55~K and (b) 12~K. Grey lines are raw data. The data was averaged by a 32~point sliding window to remove vibrational noise (blue lines), and then smoothed (red lines) to facilitate the linear fits above and below the ``elbow'' (black lines). The position of the critical field corresponds to the field at which the slope of the smoothed curve equals the average slope of the two linear fits.
All data are recorded in falling field to allow for a higher point density as the down sweep of the 65-T short-puse magnet happens on a longer time scale (90~ms) than the up sweep (9~ms)~\cite{singleton_2016}.

\section{Field-dependent masses and Lifshitz-Kosevich fits}
\label{sec:LKfits}

\begin{figure}
\begin{center}
		{\includegraphics[width=.99\columnwidth]{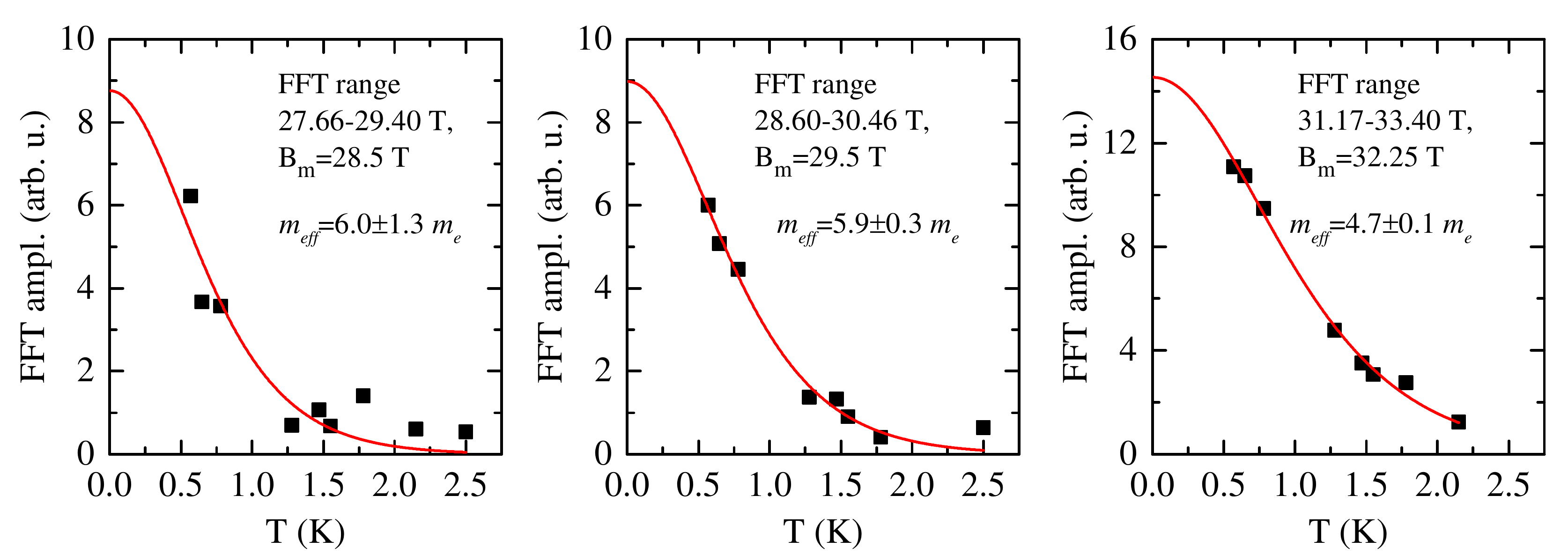}}
		\end{center}
   \caption[]{Temperature dependence of FFT peaks for three different FFT ranges and mean fields. Red lines: Fit lines obtained by the Lifshitz-Kosevich model. See labels in graphs for details on FFT ranges and effective masses.}
	\label{fig:LK_fits}
\end{figure}

In order to determine the effective masses of the quasiparticles at different mean fields $B_m$, we used small field windows that contain $\sim3.5$ oscillations corresponding to $\Delta=\unit[1/467]{T^{-1}}$ in inverse field. We have performed Fast Fourier Transforms (FFT) for around 40 (overlapping) sections of size $1/\Delta$ with different $B_m$ in the field region for which quantum oscillations had been observed, similar to \cite{rebar}. For each section a function following the Lifshitz-Kosevich (LK) model \cite{shoenberg} was fitted to the amplitudes of the FFT peaks at different temperatures. Examples for LK fits to FFT peaks at three different mean fields are shown in Fig.~\ref{fig:LK_fits} to illustrate the increase of the effective mass at lower fields. The FFT windows, mean fields $B_m$ and effective masses are given in the labels in the graphs.

\section{Quantum limit of quantum oscillations in the \texorpdfstring{$\cal{L}$}\xspace~phase}
\label{sec:QOinL}

The small size of the Fermi surface in the $\cal{L}$ phase, in combination with the transition field of the field-tuned valence transition being in the order of 10 to 50~T, raises the interesting question of whether the quantum limit from the small Fermi surface pockets in the $\cal{L}$-phase could somehow affect the phase boundary. However, the discussion below shows that the energy shifts of the $\cal{L}$-phase Landau levels do not exert a significant influence on the $\cal{L-H}$ phase boundary. 

First, the $\cal{L}$ phase is a compensated semimetal with equally populated electron and hole bands \cite{harima_2003,yan_2012} of similar mass; upward energy shifts of the electron Landau levels will be compensated by similar downward motion of the corresponding hole levels. 

Second, the quasiparticle masses in the $\cal{L}$ phase are large, as evidenced by the substantial electronic contribution to the low-temperature heat capacity \cite{bauer_2001}. Hence, the field-induced energy shifts of the hole and electron Landau levels will be very small compared to the other energy scales in the problem.

Third, the quantum limit should not be temperature dependent; also, Landau-level-based phenomena tend to become weaker as the temperature rises, due to the broadening of the Fermi-Dirac distribution function \cite{shoenberg,singleton_2001}. This is the opposite of the behavior of the phase boundary; rather than staying in place and weakening as the temperature rises, it shifts dramatically to higher fields, reaching close to 50~T by 30~K. 

Based on band-structure calculations \cite{harima_2003,yan_2012}, we estimate the quantum-oscillation frequencies for both the electron and hole pockets to be around 16~T, giving a quantum limit around $B = 2F = \unit[32]{T}$.  This is considerably higher than the transition field at low temperatures (10-20~T), suggesting that the quantum limit does not affect the valence transition significantly. 

Finally, quantum oscillations in the $\cal{L}$ phase have not been observed experimentally, supporting the proposal that the influence of the Landau levels on the quasiparticle free energy and/or density of states is small under the conditions reported here.